\documentstyle[axodraw,epsfig]{frank}

\def\d{\,{\rm d}}
\def\bcn{\begin{center}}
\def\ecn{\end{center}}
\newcommand{\tenrm}{}
\newcommand{\bkg}{background}
\newcommand{\lk}{leptoquark}
\newcommand{\bea}{\begin{eqnarray}}
\newcommand{\eea}{\end{eqnarray}}
\newcommand{\beq}{\begin{equation}}
\newcommand{\eeq}{\end{equation}}

\newcommand{\dof}{degrees of freedom}

\newcommand{\lc}{linear collider}
\newcommand{\sm}{standard model}

\newcommand{\xs}{cross section}

\newcommand{\EM}{electromagnetic}
\newcommand{\EW}{electroweak}

\newcommand{\trm}{transverse momentum}

\newcommand{\cm}{center of mass}

\newcommand{\pe}{\mbox{$e^+e^-$}}

\newcommand{\ep}{\mbox{$e^-\gamma$}}

\def\lr3{$SU(3)_L\otimes SU(3)_R$}

\def\z0{$Z^0$}
\def\Z0{$Z^0$}

\def\wrt{with respect to}

\def\cm{centre of mass}
\def\gsim{\buildrel{\lower.7ex\hbox{$>$}}\over{\lower.7ex\hbox{$\sim$}}}
\def\lsim{\buildrel{\lower.7ex\hbox{$<$}}\over{\lower.7ex\hbox{$\sim$}}}
\newenvironment{comment}[1]{}{}

\begin{document}

\begin{flushright}
MPI-PhT/95-83\\
\end{flushright}

\vskip2cm

\begin{frontmatter}
\title{Leptoquark Production in \ep\ Scattering}
\author{Frank Cuypers\thanksref{newadd}}
\address{{\tt cuypers@pss058.psi.de}\\
        Max-Planck-Institut f\"ur Physik,
        Werner-Heisenberg-Institut,
        F\"ohringer Ring 6,
        D--80805 M\"unchen,
        Germany}
\thanks[newadd]{New address as of 1 November 1995:
	Paul Scherrer Institute,
	CH-5232 Villigen PSI,
	Switzerland}
\begin{abstract}
We perform a model independent analysis
of the production of scalar and vector \lk s
in the \ep\ mode of a \lc\ of the next generation.
Since these \lk s are produced singly,
higher masses can be probed
than in other collider modes,
like \pe\ scattering.
We discuss the discovery potential
and show how polarization and angular distributions
can be used to distinguish between the different types of \lk s.
\end{abstract}
\end{frontmatter}
\clearpage

\section{Introduction}

The \sm\ of strong and \EW\ interactions,
though extremely successful in describing present day data,
is known to be plagued by a number of shortcomings.
There is no doubt that
it is merely the low energy limit
of a more profound underlying theory.
In particular,
the peculiar similarities between the quark and lepton sectors
and their miraculous anomaly cancellations
are indicative of a deeper interconnection
between these two types of particles.

Many models have been proposed
which establish a closer link between quark and lepton \dof.
They generally involve a new class of fields
which carry both lepton and baryon number
and mediate lepton-quark transitions.
Such bosons
can come in many combinations of the different quantum numbers
and are generically called \lk s.
They can,
for example,
emerge as composite scalar or vector states
of techni-fermions \cite{etc} or preons \cite{comp}.
Leptoquarks also arize naturally
as gauge vectors or Higgs scalars
in many grand unified models \cite{gut}
or superstring-derived models \cite{ss}.

In principle,
$e^-p$ scattering provides the privileged reaction for discovering \lk s.
Direct searches at HERA \cite{brw,ed}
should be able to exclude \lk s below {\em ca} 300 GeV
and a coupling to leptons and quarks above $0.03e$.
The LEP-LHC combination could of course extend these limits much further
and would also provide a powerful tool
for discriminating different \lk\ types \cite{slava}.
However,
today the most stringent limits
still originate from low-energy experiments \cite{sacha}.
These bounds could be greatly improved
at \lc s of the next generation,
though.
In particular,
the \pe\ mode is very promising
because it can abundantly pair-produce \lk s
even if their couplings to leptons and quarks are tiny \cite{br}.

In this paper
we consider the production of single \lk s
in \ep\ collisions,
with laser backscattered photons.
This reaction proceeds inevitably
via the \lk-lepton-quark coupling,
and may thus be suppressed if the latter is small.
However,
it may probe much higher \lk\ masses
than the electron-positron annihilation \cite{br}
or photon-gluon fusion \cite{ed}
processes,
which necessarily need to produce two \lk s.
Moreover,
since the \sm\ \bkg s can easily be rendered harmless
the data analysis should be exceedingly simple,
in contrast to the electron-quark fusion reaction \cite{brw}.
The angular distributions of the emerging \lk s
have complicated patterns,
which can be used advantageously
to tell apart different types of \lk s.

The study of scalar \lk\ production
in \ep\ collisions
was first discussed in Ref.~\cite{hp}.
Similarly, vector \lk s were considered in Ref.~\cite{ce}.
It was also shown \cite{nl}
that even kinematically inaccessible scalar \lk s
can also be probed this way
if their couplings to fermions are large.
For lighter scalar \lk s,
it was pointed out \cite{bln}
that substantial rates can be obtained
even with Weizs\"acker-Williams photons
in \pe\ scattering.
Because of the hadronic content of the photon,
\lk s can also be probed via electron-quark fusion
in \ep\ scattering \cite{eboli}.
It was also shown that this resolved photon contribution
may help determining \lk\ properties \cite{dg}.

In the next section
we introduce a model-independent framework
for describing a large class of \lk s \cite{brw}
and provide the analytical expressions
of the integrated \xs s
for producing the various types of \lk s.
We then shortly review the generation and properties
of the photon beam \cite{ginzburg}.
After this,
we discuss the \lk\ discovery potential
of \ep\ scattering,
which is conveniently summarized by Eq.~(\ref{osc}).
Finally,
in the last section
we discuss
the possibility of determining
the nature of the discovered \lk s
with the help of polarized beams
and at hand of their angular distributions.

\section{Cross sections}

Because of the large number of possible \lk\ types,
it is important to perform an analysis
which is as model-independent as possible.
Therefore,
to describe the \lk-lepton-quark interactions
we use the most general
$SU(3)_c \otimes SU(2)_L \otimes U(1)_Y$ invariant
effective lagrangian
of lowest dimension
which conserves lepton $(L)$ and baryon $(B)$ number.
It can be separated into two parts,
each involving
either \lk s which carry no fermion number $F=3B+L=0$,
or \lk s with fermion number $F=2$ \cite{brw}:
\begin{eqnarray}
{\cal L}_{F=0} \quad =
& &
\bigl(
  h_{2L} \bar{u}_R \ell_L
+ h_{2R} \bar{q}_L i \sigma_2 e_R
\bigr) R_2
+ \tilde{h}_{2L} \bar{d}_R \ell_L \tilde{R}_2
\\
& + &
\bigl(
  h_{1L} \bar{q}_L \gamma^\mu \ell_L
+ h_{1R} \bar{d}_R \gamma^\mu e_R
\bigr) U_{1 \mu}
+ \tilde{h}_{1R} \bar{u}_R \gamma^\mu e_R \tilde{U}_{1 \mu}
\nonumber\\
& + &
h_{3L} \bar{q}_L \mbox{\boldmath$\sigma$} \gamma^\mu \ell_L
\mbox{\boldmath$U$}_{3 \mu}
\nonumber\\
& + & \mbox{ h.c.}
\nonumber\\
\nonumber\\
{\cal L}_{F=2} \quad =
& &
\bigl(
  g_{1L} \bar{q}^c_L i\sigma_2 \ell_L
+ g_{1R} \bar{u}^c_R e_R
\bigr) S_1
+ \tilde{g}_{1R} \bar{d}^c_R e_R \tilde{S}_1
\\
& + &
g_{3L} \bar{q}^c_L i\sigma_2\mbox{\boldmath$\sigma$} \ell_L
\mbox{\boldmath$S$}_3
\nonumber\\
& + &
\bigl(
  g_{2L} \bar{d}^c_R \gamma^\mu \ell_L
+ g_{2R} \bar{q}^c_L \gamma^\mu e_R
\bigr) V_{2 \mu}
+ \tilde{g}_{2L} \bar{u}^c_R \gamma^\mu \ell_L \tilde{V}_{2 \mu}
\nonumber\\
& + & \mbox{ h.c.}
\nonumber
\end{eqnarray}
where the \boldmath$\sigma$'s\unboldmath\ are Pauli matrices,
while $q_L$ and $\ell_L$ are the $SU(2)_L$ quark and lepton doublets
and $u_R$, $d_R$, $\ell_R$ are the corresponding singlets.
The subscripts of the \lk s
indicates the size of the $SU(2)_L$ representation
they belong to.
The $R$- and $S$-type \lk s are spacetime scalars,
whereas the $U$ and $V$ are vectors.
Family and colour indices are implicit.

Since all these \lk s carry an electric charge,
they must also couple to the photon.
These interactions are described by the kinetic lagrangians
for scalar and and vector bosons \cite{ed}
\begin{eqnarray}
{\cal L}_{J=0}
& = &
\sum_{\rm{scalars}} \quad
\left(D_\mu\Phi\right)^{\dag} \left(D^\mu\Phi\right)
-
m^2 \Phi^{\dag}\Phi
\\
{\cal L}_{J=1}
& = &
\sum_{\rm{vectors}} \quad
-{1\over2} \left(D_\mu\Phi^\nu-D_\nu\Phi^\mu\right)^{\dag}
           \left(D^\mu\Phi_\nu-D^\nu\Phi_\mu\right)
+
m^2 \Phi_\mu^{\dag}\Phi^\nu
\label{comp}
\end{eqnarray}
with the covariant derivative
\beq
D_\mu = \partial_\mu - ieQA_\mu
\ ,
\eeq
where $\Phi$ and $A$ are the \lk\ and photon fields,
$m$ and $Q$ are the \lk\ mass and \EM\ charge
and $e$ is the \EM\ coupling constant.
This lagrangian describes the minimal vector boson coupling,
typical of a composite \lk.
If, however,
the vector \lk s are gauge bosons,
an extra Yang-Mills piece has to be added
in order to maintain gauge invariance \cite{ed}:
\beq
{\cal L}_{G} =
\sum_{\rm{vectors}} \quad
-ie ~\Phi_\mu^{\dag}\Phi_\nu \left(\partial^\mu A^\nu - \partial^\nu
A^\mu\right)
\ .\label{ym}
\eeq
\begin{comment}{
If this piece is not included,
tree-level unitarity is bound to be lost.
This, however,
is expected,
since the effective lagrangian (\ref{comp})
is no longer valid at energies
of the order of the compositeness scale.
At those higher energies
the terms which decouple at lower energies become relevant
and the full (gauge) theory from which (\ref{comp}) was derived
has to be considered.
}\end{comment}

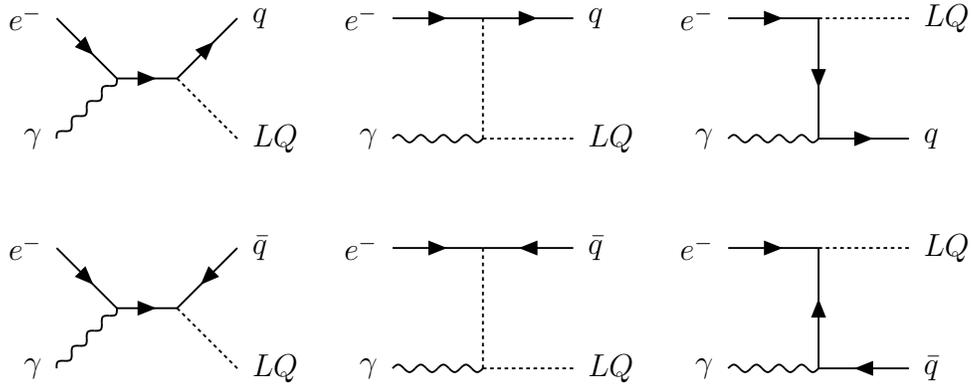
\begin{figure}[htb]
{\unitlength.5mm\SetScale{1.4185}

\begin{picture}(70,40)(-30,16)
\Text(-4,32)[rc]{$e^-$}
\Text(-4,00)[rc]{$\gamma$}
\ArrowLine(00,32)(16,16)
\ArrowLine(16,16)(32,16)
\ArrowLine(32,16)(48,32)
\DashLine(32,16)(48,00){1}
\Photon(00,00)(16,16){1}{4}
\Text(52,32)[lc]{$q$}
\Text(52,00)[lc]{$LQ$}
\end{picture}
\qquad
\begin{picture}(70,40)(-30,16)
\Text(-4,32)[rc]{$e^-$}
\Text(-4,00)[rc]{$\gamma$}
\ArrowLine(00,32)(24,32)
\ArrowLine(24,32)(48,32)
\DashLine(24,32)(24,00){1}
\DashLine(24,00)(48,00){1}
\Photon(00,00)(24,00){1}{4}
\Text(52,32)[lc]{$q$}
\Text(52,00)[lc]{$LQ$}
\end{picture}
\qquad
\begin{picture}(70,40)(-30,16)
\Text(-4,32)[rc]{$e^-$}
\Text(-4,00)[rc]{$\gamma$}
\ArrowLine(00,32)(24,32)
\ArrowLine(24,32)(24,00)
\ArrowLine(24,00)(48,00)
\DashLine(24,32)(48,32){1}
\Photon(00,00)(24,00){1}{4}
\Text(52,32)[lc]{$LQ$}
\Text(52,00)[lc]{$q$}
\end{picture}

\bigskip\bigskip

\begin{picture}(70,40)(-30,16)
\Text(-4,32)[rc]{$e^-$}
\Text(-4,00)[rc]{$\gamma$}
\ArrowLine(00,32)(16,16)
\ArrowLine(16,16)(32,16)
\ArrowLine(48,32)(32,16)
\DashLine(32,16)(48,00){1}
\Photon(00,00)(16,16){1}{4}
\Text(52,32)[lc]{$\bar q$}
\Text(52,00)[lc]{$LQ$}
\end{picture}
\qquad
\begin{picture}(70,40)(-30,16)
\Text(-4,32)[rc]{$e^-$}
\Text(-4,00)[rc]{$\gamma$}
\ArrowLine(00,32)(24,32)
\ArrowLine(48,32)(24,32)
\DashLine(24,32)(24,00){1}
\DashLine(24,00)(48,00){1}
\Photon(00,00)(24,00){1}{4}
\Text(52,32)[lc]{$\bar q$}
\Text(52,00)[lc]{$LQ$}
\end{picture}
\qquad
\begin{picture}(70,40)(-30,16)
\Text(-4,32)[rc]{$e^-$}
\Text(-4,00)[rc]{$\gamma$}
\ArrowLine(00,32)(24,32)
\ArrowLine(24,00)(24,32)
\ArrowLine(48,00)(24,00)
\DashLine(24,32)(48,32){1}
\Photon(00,00)(24,00){1}{4}
\Text(52,32)[lc]{$LQ$}
\Text(52,00)[lc]{$\bar q$}
\end{picture}

\bigskip\bigskip

}
\bigskip
\caption{
  Lowest order Feynman diagrams for the production of
  $F=0$ (upper graphs) and
  $F=2$ (lower graphs) \lk s.
}
\label{feyn}
\end{figure}

Ignoring the resolved photon contributions,
the following \lk\ reactions are possible to lowest order in \ep\ collisions:
\begin{eqnarray*}
F=0:~\left\{
\begin{array}{rl}
  ~J=0:~\left\{
  \begin{array}{@{~}r@{~}c@{~}l}
    e^-_R ~ \gamma & \to
    &  u_L ~ R_2^{-5/3} \\
    && d_L ~ R_2^{-2/3} \\
  \end{array}
  \qquad
  \begin{array}{@{~}r@{~}c@{~}l}
    e^-_L ~ \gamma & \to
    &  u_R ~ R_2^{-5/3} \\
    && d_R ~ \tilde R_2^{-2/3} \\
  \end{array}
  \right.
  \\\\
  ~J=1:~\left\{
  \begin{array}{@{~}r@{~}c@{~}l}
    e^-_R ~ \gamma & \to
    & u_R ~ \tilde U_1^{-5/3} \\
    && d_R ~ U_1^{-2/3} \\\\
  \end{array}
  \qquad
  \begin{array}{@{~}r@{~}c@{~}l}
    e^-_L ~ \gamma & \to
    &  u_L ~ U_3^{-5/3} \\
    && d_L ~ U_1^{-2/3} \\
    && d_L ~ U_3^{-2/3} \\
  \end{array}
  \right.
\end{array}
\right.
\label{f0}
\end{eqnarray*}
\begin{eqnarray*}
F=2:~\left\{
\begin{array}{rl}
  ~J=0:~\left\{
  \begin{array}{@{~}r@{~}c@{~}l}
    e^-_R ~ \gamma & \to
    &  \bar u_R ~ S_1^{-1/3} \\
    && \bar d_R ~ \tilde S_1^{-4/3} \\\\
  \end{array}
  \qquad
  \begin{array}{@{~}r@{~}c@{~}l}
    e^-_L ~ \gamma & \to
    &  \bar u_L ~ S_1^{-1/3} \\
    && \bar u_L ~ S_3^{-1/3} \\
    && \bar d_L ~ S_3^{-4/3} \\
  \end{array}
  \right.
  \\\\
  ~J=1:~\left\{
  \begin{array}{@{~}r@{~}c@{~}l}
    e^-_R ~ \gamma & \to
    & \bar u_L ~ V_2^{-1/3} \\
    && \bar d_L ~ V_2^{-4/3} \\
  \end{array}
  \qquad
  \begin{array}{@{~}r@{~}c@{~}l}
    e^-_L ~ \gamma & \to
    & \bar u_R ~ \tilde V_2^{-1/3} \\
    && \bar d_R ~ V_2^{-4/3} \\
  \end{array}
  \right.
\end{array}
\right.
\label{f2}
\end{eqnarray*}
The superscripts of the \lk s indicate their \EM\ charge.
The typical $s$-, $t$- and $u$-channel Feynman diagrams
for the $F=0$ and $F=2$ \lk\ production
are shown in Fig.~\ref{feyn}.
The \xs s, though,
do not depend explicitly on this quantum number.

There are 24 different types of processes,
depending on
whether the produced \lk\ is a scalar, vector or gauge boson,
whether it couples to right- or left-handed leptons
and what is its \EM\ charge ($Q=-1/3,-2/3,-4/3,-5/3$).
The differential \xs s are lengthy
and we do not report them here.
We agree with the unpolarized expressions
reported in Refs~\cite{hp,ce,nl}\footnote{
  The colour factor is not explicitly stated in Refs~\cite{hp,nl}.
},
though.

Let us define
\beq
x = {m^2 \over s}
\ ,\label{x}
\eeq
where $m$ is the \lk\ mass and $s$ is the \cm\ energy squared.
We also use the generic \lk\ coupling $\lambda$ to the electrons and quarks,
and denote the electron and photon polarizations
$P_e$ and $P_\gamma$.
We find for the integrated scalar \xs s
\bea
 \makebox[0cm][l]{\hskip-5mm$\displaystyle
   \sigma(J=0)
   \quad =
   \quad {3\pi\alpha^2\over2s}
   \quad \left({\lambda\over e}\right)^2
   \quad {1 \pm P_e \over 2}
   \quad \times$}
\label{s0}\\
& \biggl\{\quad &
\left(-\left(3+4Q\right) + \left(7+8Q+8Q^2\right)x\right)
\quad (1-x)
\nonumber\\
&&
+4Q\left( Q - \left(2+Q\right)x \right)
\quad x \ln x
\nonumber\\
&&
-2\left(1+Q\right)^2\left( 1 - 2x + 2x^2 \right)
\quad \ln{m_q^2/s\over\left(1-x\right)^2}
\nonumber\\
& \pm~P_\gamma~\Bigl[~ &
\left(-\left(7+12Q+4Q^2\right) + 3x\right)
\quad (1-x)
\nonumber\\
&&
+4Q^2
\quad x \ln x
\nonumber\\
&&
-2\left(1+Q\right)^2\left( 1 - 2x \right)
\quad \ln{m_q^2/s\over\left(1-x\right)^2}
\quad \Bigr] \quad \biggr\}
\nonumber~.
\eea
The unpolarized part
($P_e=P_\gamma$=0)
of this expression
agrees with the result reported in Ref.~\cite{bln},
up to the colour factor,
which we include explicitly here.
The vector \xs s are
\bea
 \makebox[0cm][l]{\hskip-5mm$\displaystyle
   \sigma(J=1)
   \quad =
   \quad {3\pi\alpha^2\over8m^2}
   \quad \left({\lambda\over e}\right)^2
   \quad {1 \pm P_e \over 2}
   \quad \times$}
\label{c0}\\
& \biggl\{\quad &
\left(Q^2 + \left(8-16Q-Q^2\right)x + 8\left(7+8Q+8Q^2\right)x^2\right)
\quad (1-x)
\nonumber\\
&&
-4Q\left( Q + \left(8+3Q\right)x - 8Qx^2 + 8\left(2+Q\right)x^3 \right)
\quad \ln x
\nonumber\\
&&
-16\left(1+Q\right)^2\left( 1 - 2x + 2x^2 \right)
\quad x \ln{m_q^2/s\over\left(1-x\right)^2}
\nonumber\\
& \pm~P_\gamma~\Bigl[~ &
\left(- 3Q^2 + \left(40+48Q+63Q^2\right)x + 24x^2\right)
\quad (1-x)
\nonumber\\
&&
-4Q\left( -3Q + 8\left(3+Q\right)x \right)
\quad x \ln x
\nonumber\\
&&
+16\left(1+Q\right)^2\left( 1 - 2x \right)
\quad x \ln{m_q^2/s\over\left(1-x\right)^2}
\quad \Bigr] \quad \biggr\}
\nonumber~.
\eea
Finally,
if the vector is a gauge field,
we have
\bea
 \makebox[0cm][l]{\hskip-5mm$\displaystyle
   \sigma^{\rm YM}(J=1)
   \quad =
   \quad {3\pi\alpha^2\over m^2}
   \quad \left({\lambda\over e}\right)^2
   \quad {1 \pm P_e \over 2}
   \quad \times$}
\label{v0}\\
& \biggl\{\quad &
\left(4Q^2 + \left(1-4Q\right)x + \left(7+8Q+8Q^2\right)x^2\right)
\quad (1-x)
\nonumber\\
&&
-4Q\left( 2 - Qx + \left(2+Q\right)x^2 \right)
\quad x \ln x
\nonumber\\
&&
-2\left(1+Q\right)^2\left( 1 - 2x + 2x^2 \right)
\quad x \ln{m_q^2/s\over\left(1-x\right)^2}
\nonumber\\
& \pm~P_\gamma~\Bigl[~ &
\left(\left(5+4Q+12Q^2\right) + 3x\right) x
\quad (1-x)
\nonumber\\
&&
+4Q\left( Q - \left(4+Q\right)x \right)
\quad x \ln x
\nonumber\\
&&
+2\left(1+Q\right)^2\left( 1 - 2x \right)
\quad x \ln{m_q^2/s\over\left(1-x\right)^2}
\quad \Bigr] \quad \biggr\}
\nonumber~.
\eea
The electron and quark masses have been set equal to zero everywhere
to derive Eqs~(\ref{s0}-\ref{v0}),
except in the squared $u$-channel matrix elements.
It is essential to perform the calculation of these terms
with a finite quark mass,
because it regulates the singularity
which occurs when a \lk\ is emitted in the direction of the incoming electron.
The approximation
$\ln m_q^2/s/(1-x)^2$
we have written
for this $u$-channel logarithm
stays more than accurate
up to within a few quark masses from the threshold.
It is straightforward to compute the full expression \cite{bln}.
Moreover,
the integration in the backward direction
introduces non-vanishing contributions of the order
${\cal O}(m_q^2/m_q^2)$
to the non-logarithmic parts of the \xs s (\ref{s0}-\ref{v0})\footnote{
  I am very much indebted to David London and H\'el\`ene Nadeau
  for pointing this out to me.
}.

The threshold behaviour of the \xs s,
around $x=1$,
mainly depends on the $u$-channel singularity
and on the electron and photon relative polarizations:
\beq
x=1: \left\{ \quad
\begin{array}{l}
  \displaystyle{\sigma(J=0) \quad \propto \quad ~(1+Q)^2 
    \quad (1 \mp P_\gamma) (1 \pm P_e) }
  \\\\
  \displaystyle{\sigma(J=1) \quad \propto \quad 2(1+Q)^2 
    \quad (1 \pm P_\gamma) (1 \pm P_e) }
\end{array}
\right.
\label{thresh}
\eeq
The scalar \xs\ quickly drops to zero
for like-sign initial state electrons and photons polarizations,
whereas the vector \xs\ is suppressed
when these polarizations have opposite signs.
For equal couplings,
the vector threshold \xs s are twice as intense as the scalar ones.

In the asymptotic region,
for $x=0$,
the $J=0$ and $J=1$ \lk s also display very different behaviours.
Whereas the scalar \xs s decrease like
$1/s$,
the vector \xs s eventually increase like
$\ln s$.
If the vectors are gauge fields, though,
their \xs s saturate for large values of $s$.

\section{Photon Beams}

The \xs s (\ref{s0}-\ref{v0})
still have to be folded
with a realistic Compton backscattered photon spectrum \cite{ginzburg}:
\def\d{\,{\rm d}}
\beq
\sigma(s) = \int_{x_{\rm min}}^{x_{\rm max}} \d x
{\d n(x) \over \d x} \sigma(xs)
\ ,\label{fold}
\eeq
where the probability density of a photon to have the energy fraction
$x = E_\gamma/E_{\rm beam}$
is given by
\bea
{\d n(x) \over \d x} =
\frac{1}{\cal N}
& \Biggl\{ & 1 - x + \frac{1}{1 - x} - \frac{4 x }{z ( 1 - x) }
                           + \frac{4 x^2 } {z^2 (1 - x)^2} \nonumber \\
& + & P_{\rm beam} P_{\rm laser} \frac{x (2 - x)}{1 - x}
                       \left[ \frac{2 x} {z (1 - x)} - 1 \right]
\Biggr\}
\ , \label{spec}
\eea
where
\bea
{\cal N}
& = &
        \frac{z^3 + 18 z^2 + 24 z + 8}{2 z (z + 1)^2 }
        + \left( 1 - \frac{4}{z} - \frac{8}{z^2} \right) \ln (1 + z)
\\
& + &
        P_{\rm beam} P_{\rm laser} \left[ 2 - \frac{z^2}{(z + 1)^2}
        - \left( 1 + \frac{2}{z} \right) \ln (1 + z) \right]
\ ,\nonumber
\eea
\beq
z = \frac{4 E_{\rm beam} E_{\rm laser} }{m_e^2}
\ ,
\eeq
$m_e$ being the mass of the electron,

By design,
the energy fraction $x$ of the photons is limited from above to
\beq
x_{\rm max} = \frac{ 2+2\sqrt{2}}{ 3+2\sqrt{2}} \approx 0.8284
\label{xmax}
\eeq
in order to prevent electron-positron pair-production
from photon rescattering.
In practice,
it is also limited from below,
because only the harder photons are produced at a small angle
\wrt\ the beam-pipe
according to
\beq
\theta_\gamma (x) \simeq {m_e \over E_{\rm beam}} \sqrt{{z \over x}-z-1}
\ .
\eeq
The softer photons are emitted at such large angles
that they are bound to miss the opposite
highly collimated electron beam.
Assuming a conversion distance of {\em ca} 5 cm
and a beam size of 500 nm diameter,
the lower bound we adopt for the photon energy fraction is
\beq
x_{\rm min} = .5
\ . \label{xmin}
\eeq
Of course,
if the beams are very flat,
this cut-off will be somewhat softened out.
However,
in the threshold region is has no effect.
When it sets in,
at $s=m^2/x_{\rm min}=2m^2$,
it is visible on the plots of Fig.~\ref{eny}
as a slight kink in the slopes.

The polarization of the backscattered photons is given by
\beq
P_\gamma (x) = \frac{P_{\rm laser} \zeta (2 - 2 x + x^2)  + P_{\rm beam} x ( 1
+ \zeta^2)}
               { (1 -x ) (2 - 2 x + x^2) - 4 x ( z - z x - x)/z^2
                         - P_{\rm beam} P_{\rm laser} \zeta x ( 2 - x) }
\ , \label{polar}
\eeq
with
\beq
\zeta = 1 - x ( 1 + 1/z)
\ . \label{e103}
\eeq

All these features of the Compton backscattered photon beams
are displayed in Fig.~\ref{photon}.
In particular,
it appears that
if the signs
of the polarizations of the laser and the Compton scattered electron beams
are chosen to be opposite,
we obtain the most intense, hard and highly polarized photon beam.
We adopt this choice throughout the rest of this paper.

\section{Leptoquark Discovery Limits}

To present our results,
we have chosen to work with a quark mass in the $u$-channel propagator
$m_q=10$ MeV.
Since the dependence of the \xs s on this mass is only logarithmic,
the error does not exceed a few \%\
and is mainly confined to the immediate threshold region.
We also assumed both the unaltered and the Compton scattered electron beams
to be 90\%\ polarized.
This should be fairly easy to realize at a \lc\ of the next generation.

In Fig.~\ref{eny}
we have displayed the behaviours of the \xs s (\ref{s0}-\ref{v0})
as functions of the collider \cm\ energy.
For the purpose of these plots,
we have set the \lk-electron-quark coupling
equal to the \EM\ coupling constant
$\lambda=e$.

In general,
\lk s which are produced in the reactions (\ref{s0}-\ref{v0}),
will decay into a charged lepton and a jet
with a substantial branching ratio.
If the \lk s are bound to a single generation,
the decay lepton is an electron.
Around threshold,
the $u$-channel pole is dominant,
so the quarks and \lk s are mostly produced at very small angles
from the beampipe.
Since the \lk s are almost at rest,
most of the electron and quark into which they decay
are emitted at large angles.
The \lk\ signature is thus
a low \trm\ (or even invisible) jet,
and a high \trm\ electron and jet pair
whose invariant mass is closely centered around the \lk\ mass.
Away from threshold,
the $u$-channel is no longer dominant,
and the \lk\ signature becomes
a high \trm\ electron and pair of jets,
where the electron and one of the jets
have a combined invariant mass close to the \lk\ mass.

If all jet pairs with an invariant mass around the $Z^0$ mass are cut out,
there is no $Z^0$ Compton scattering background.
The next order background which then subsists is
the quark photoproduction process
$\ep \to e^-q\bar q$.
This background, though,
will not show any peak in the electron-jet invariant mass distribution
at the \lk\ mass,
and can moreover be drastically reduced
by requiring a minimum \trm\ for the electron.
If one allows for \lk s
which couple equally well to different generations,
the final state can contain a muon instead of an electron.
In this case, of course,
there is no \sm\ background at all.

\begin{figure}[htb]
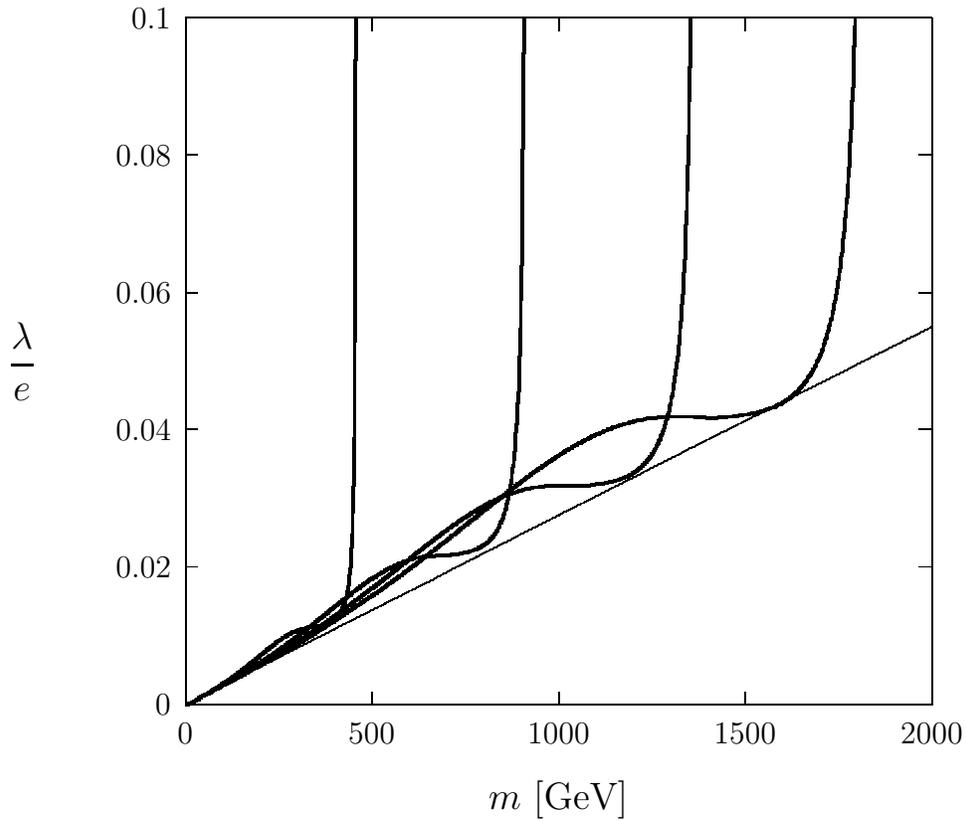

\input lim.tex
\bigskip
\caption{
  Loci of $\sigma(J=1,Q=-5/3)=1$ fb
  as a function of the \lk\ mass and coupling to fermions.
  The polarizations are $|P_e|=.9$ and $\langle P_eP_\gamma \rangle > 0$.
  The collider energies are from left to right .5, 1, 1.5 and 2 TeV.
  The thinner osculating line is given by Eq.~(\protect\ref{osc}).
}
\label{lim}
\end{figure}

To estimate the \lk\ discovery potential
of \ep\ collisions,
we have plotted in Fig.~\ref{lim}
the boundary in the $(m_{LQ},\lambda/e)$ plane,
below which the \xs\
$\sigma(J=1,Q=-5/3)$
in Eq.~(\ref{v0})
yields less than 10 events.
For this we consider four different collider energies
and assume 10 fb$^{-1}$ of accumulated luminosity.
In general,
as can be inferred from Eq.~(\ref{thresh}),
these curves are closely osculated by the relation
\beq
{\lambda \over e} \quad = \quad 0.03
\quad {m/\mbox{TeV} \over \sqrt{{\cal L}/{\rm fb}^{-1}}}
\quad \sqrt{{n \over (J+1) (1+Q)^2 }}
\qquad
\left(
  m \le .8\sqrt{s_{ee}}
\right)
\ ,\label{osc}
\eeq
where
$\lambda$ is the \lk's coupling to leptons and quarks,
$m$ its mass,
$Q$ its charge,
$J$ its spin,
$n$ is the required number of events and
$\cal L$ is the available luminosity.
This scaling relation
provides a convenient means to gauge
the \lk\ discovery potential of \ep\ scattering.
It is valid for \lk\ masses short off 80\%\ of the collider energy
and assumes
the electron beams to be 90\%\ polarized
($|P_e|=.9$)
while the chirality of the fully polarized photon beam is chosen
such as to enhance the threshold \xs\
({\em cf.} Eq.~(\ref{thresh})).

In comparison,
the best bounds on the \lk\ couplings
which have been derived indirectly from low energy data \cite{sacha}
are no better than
\beq
{\lambda \over e} \quad \ge \quad m/\mbox{TeV}
\ ,
\eeq
for \lk\ interactions involving only the first generation.
Similar bounds on couplings involving higher generations
are even poorer.

\section{Leptoquark-Type Discrimination}

If a \lk\ is discovered someday,
it is interesting to determine its nature.
In principle,
\ep\ scattering may discriminate between the 24
combinations of the quantum numbers
$J$, $Q$ and $P$,
where the latter is the chirality of the electron to which the \lk\ couples.
For the case $J=1$ we make the distinction between
gauge and non-gauge \lk s.

\begin{figure}[htb]
{\large$\sigma$ [pb]}
\vskip55mm
\epsfig{file=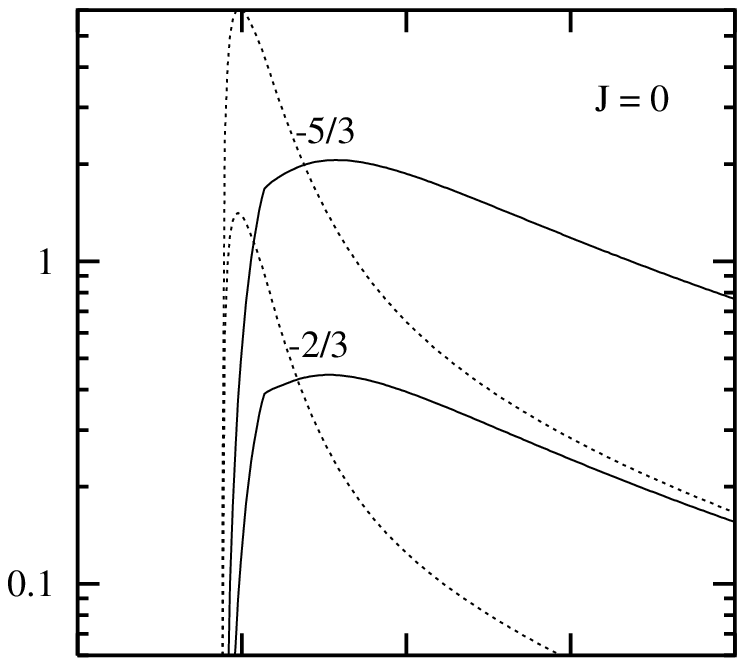,bbllx=4cm,bblly=4.2cm}
\epsfig{file=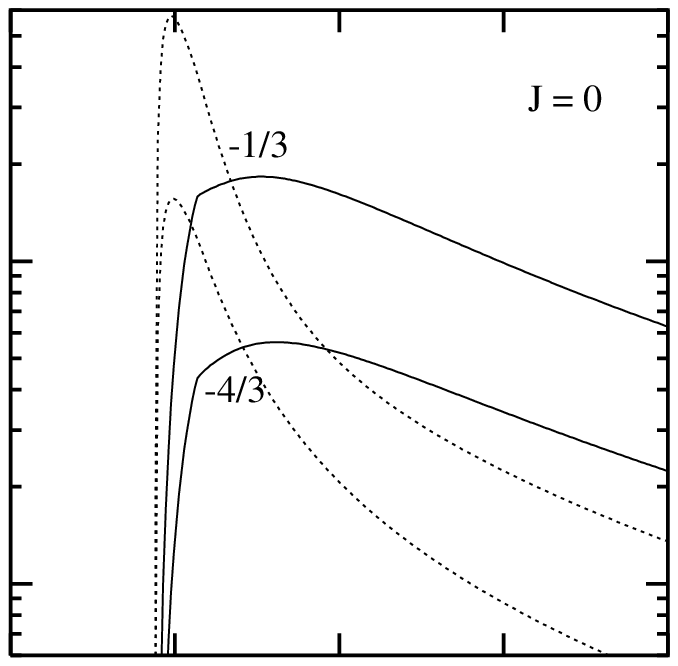,bbllx=-7cm,bblly=4.2cm}
\vskip65mm
\epsfig{file=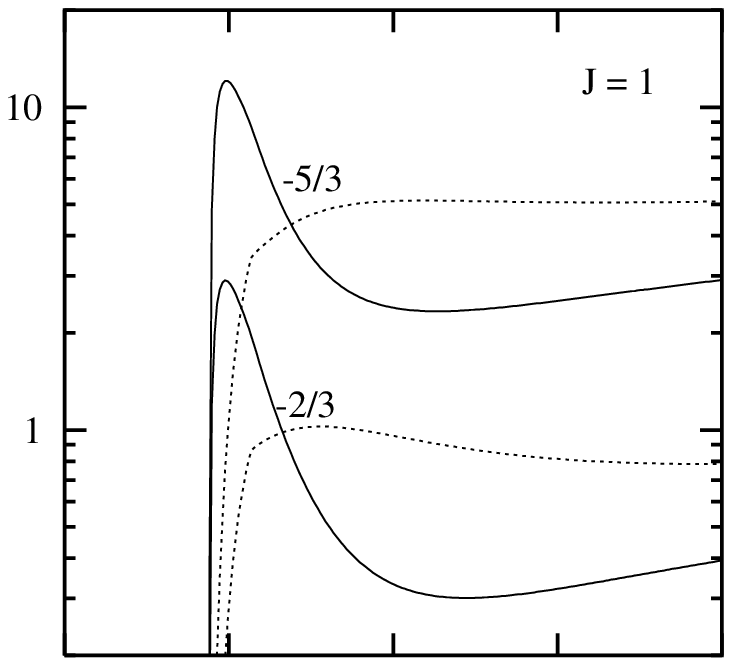,bbllx=4cm,bblly=4.2cm}
\epsfig{file=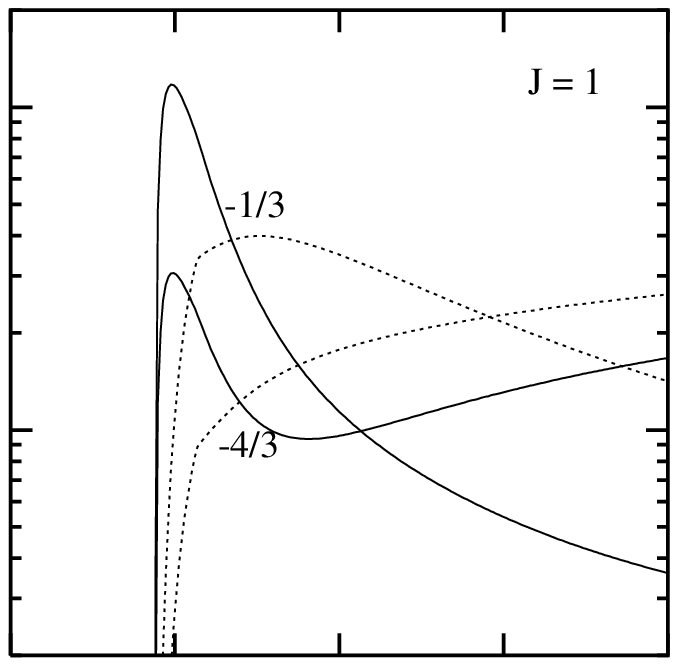,bbllx=-7cm,bblly=4.2cm}
\vskip65mm
\epsfig{file=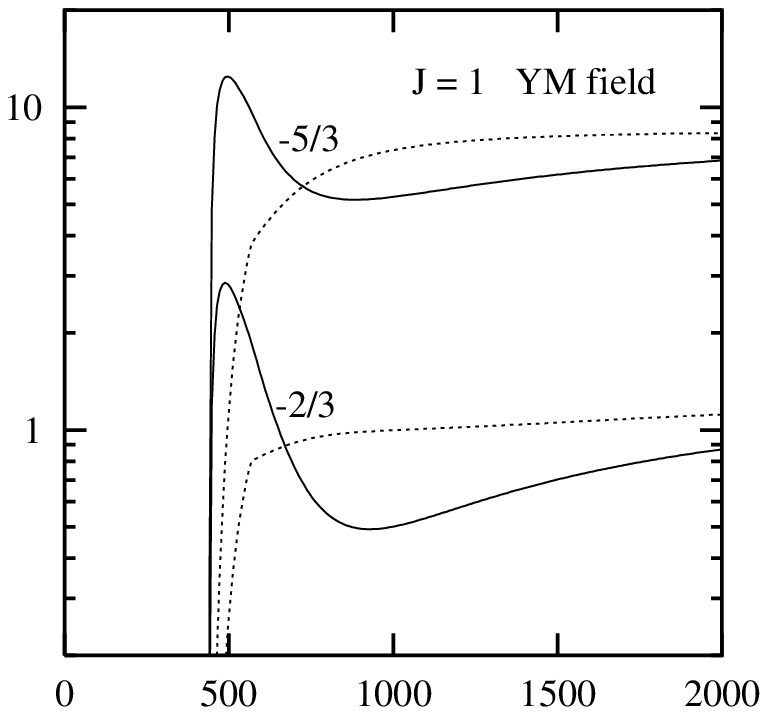,bbllx=4cm,bblly=4.2cm}
\epsfig{file=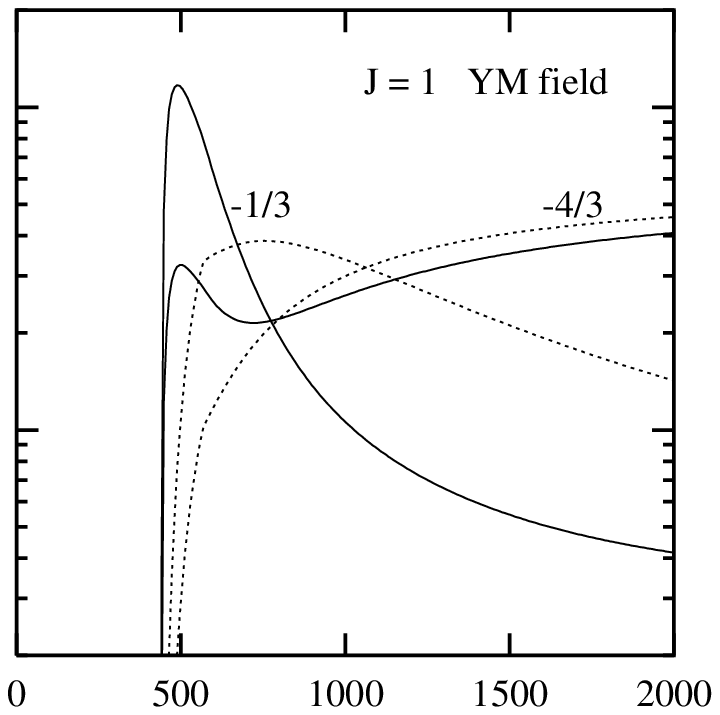,bbllx=-7cm,bblly=4.2cm}
\vskip20mm
\hskip32mm{\large$\sqrt{s}$ [GeV]}\hskip57mm{\large$\sqrt{s}$ [GeV]}
\vskip0mm
\caption{
  Dependence of the \lk\ production \xs\
  on the collider energy.
  The \lk\ mass is 400 GeV.
  The electron beams are 90\%\ polarized ($|P_e|=.9$).
  The polarization of the photons depends on their energy fraction
  and is given by Eq.~(\protect\ref{polar}).
  The solid curves correspond to the choice
  of photon and electron polarizations
  which yields $\langle P_eP_\gamma \rangle > 0$,
  whereas the dotted curves are for $\langle P_eP_\gamma \rangle < 0$.
  The charges of the \lk s are indicated next to the corresponding curves.
}
\label{eny}
\end{figure}

It is of course trivial to determine $P$
by switching the polarization of the electron beam.
Similarly,
it is almost as easy to distinguish scalars from vectors.
Indeed,
as can be inferred from Eqs~(\ref{thresh})
and from Fig.~\ref{eny},
all threshold \xs s are very sensitive
to the relative electron and photon polarizations.
Since this effect works in opposite directions
for $J=0$ and $J=1$,
a simple photon polarization flip upon discovery
should suffice to determine
the spin of the discovered \lk s.

Determining the other quantum numbers,
unfortunately,
is not as easy.
In practice this task is not facilitated
by the fact all the different models involving \lk s
predict very different values for their couplings to leptons and quarks,
if at all.
Basically,
we have no idea what to expect.
Fig.~\ref{eny} indicates that
it may be possible to discriminate some of the different vector \lk s,
by combining the information gathered from
polarization flips and an energy scan.
It will, however,
be exceedingly difficult to differentiate for instance the different
scalar \lk s from eachother
using the information from total \xs s alone.

Much more discriminating power can be obtained
from the differential distributions,
though.
We do not report their long analytical forms here.
As it turns out,
the interferences between the different channels
result into rather complex angular dependences of the \xs s.
There are even radiation zeros
for the reactions involving
scalar and Yang-Mills \lk s of charges -1/3 and -2/3.
Note that in the cross-channel reaction
$e^-q \to LQ\gamma$
these radiation zeros occur
for the scalar and Yang-Mills \lk s of charges -4/3 and -5/3  \cite{slava}.

\begin{figure}[htb]
{\large$\displaystyle{1\over\sigma}{\d\sigma \over \d\theta}$}
\hskip2.cm
\raisebox{-5mm}{\large$\displaystyle\langle P_eP_\gamma \rangle > 0$}
\hskip4.8cm
\raisebox{-5mm}{\large$\displaystyle\langle P_eP_\gamma \rangle < 0$}
\vskip-2mm
\renewcommand{\arraystretch}{3.2}{\normalsize$\begin{array}[b]{@{}r@{}}10^{-2}\\10^{-3}\\10^{-4}\\10^{-5}\end{array}$}\renewcommand{\arraystretch}{1}\hskip-3mm
\epsfig{file=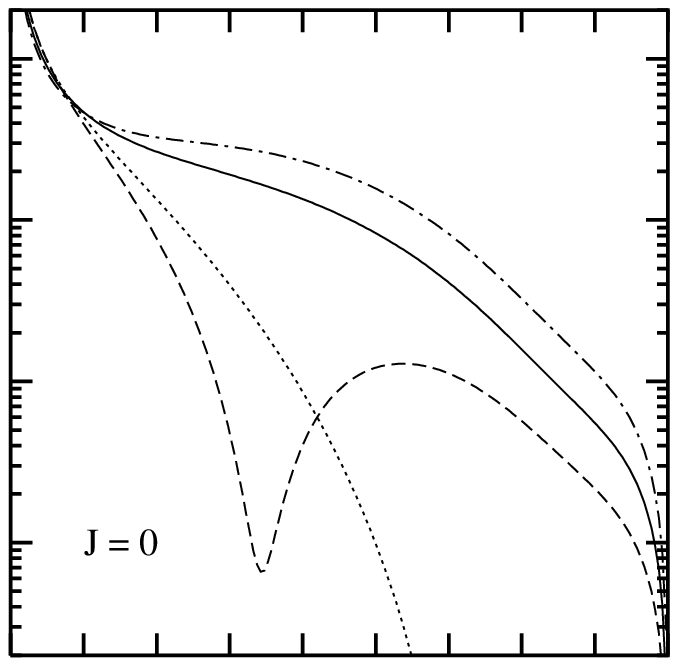,bbllx=4cm,bblly=4.2cm}
\epsfig{file=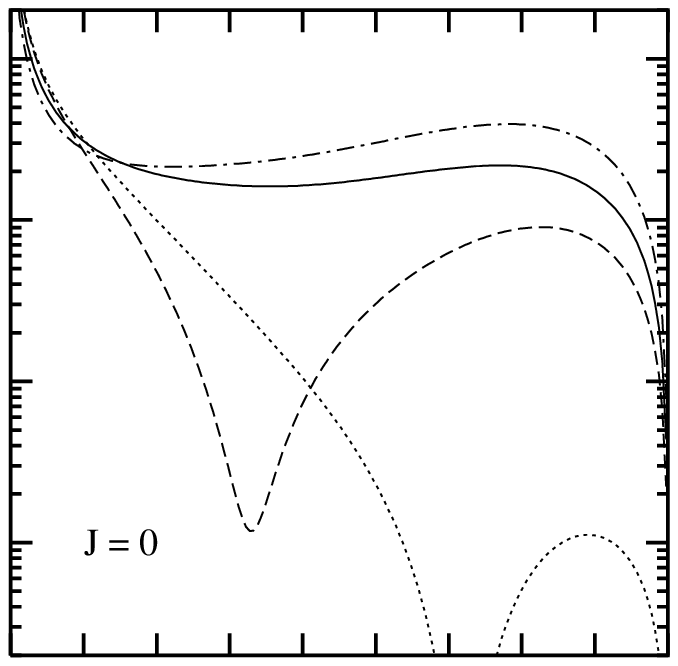,bbllx=-7cm,bblly=4.2cm}
\vskip5mm
\renewcommand{\arraystretch}{3.2}{\normalsize$\begin{array}[b]{@{}r@{}}10^{-2}\\10^{-3}\\10^{-4}\\10^{-5}\end{array}$}\renewcommand{\arraystretch}{1}\hskip-3mm
\epsfig{file=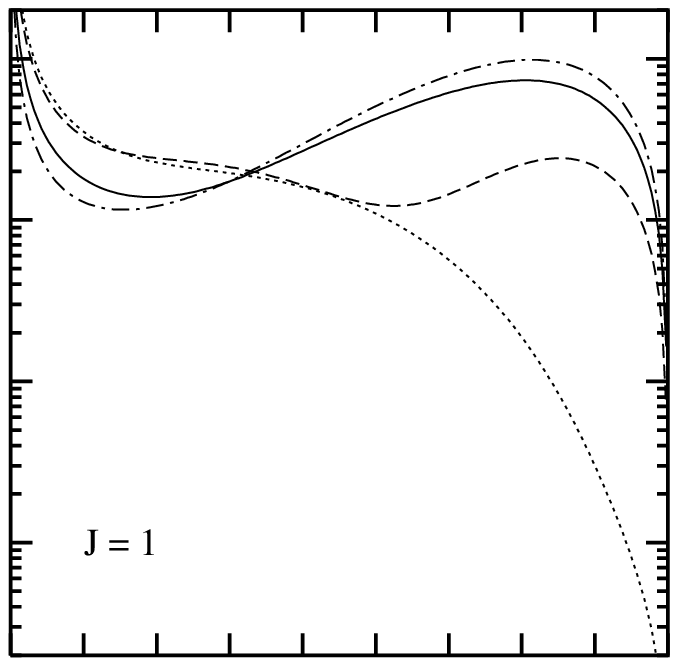,bbllx=4cm,bblly=4.2cm}
\epsfig{file=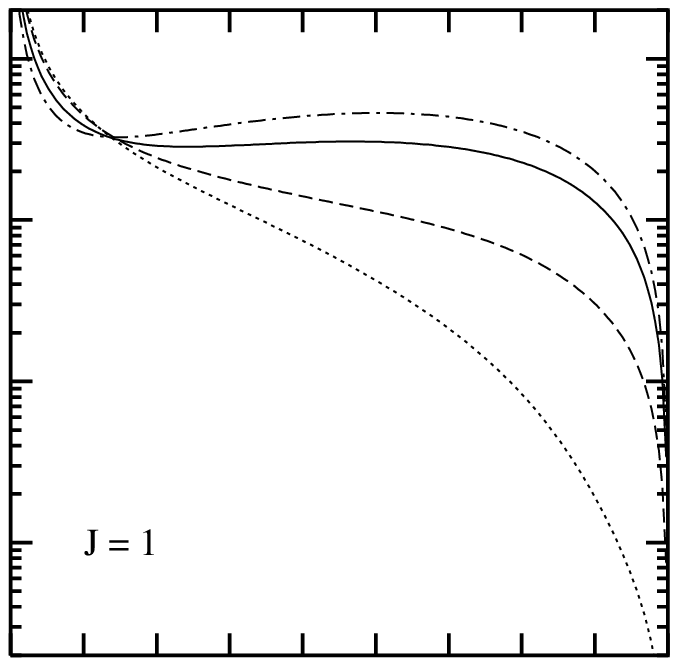,bbllx=-7cm,bblly=4.2cm}
\vskip5mm
\renewcommand{\arraystretch}{3.2}{\normalsize$\begin{array}[b]{@{}r@{}}10^{-2}\\10^{-3}\\10^{-4}\\10^{-5}\end{array}$}\renewcommand{\arraystretch}{1}\hskip-3mm
\epsfig{file=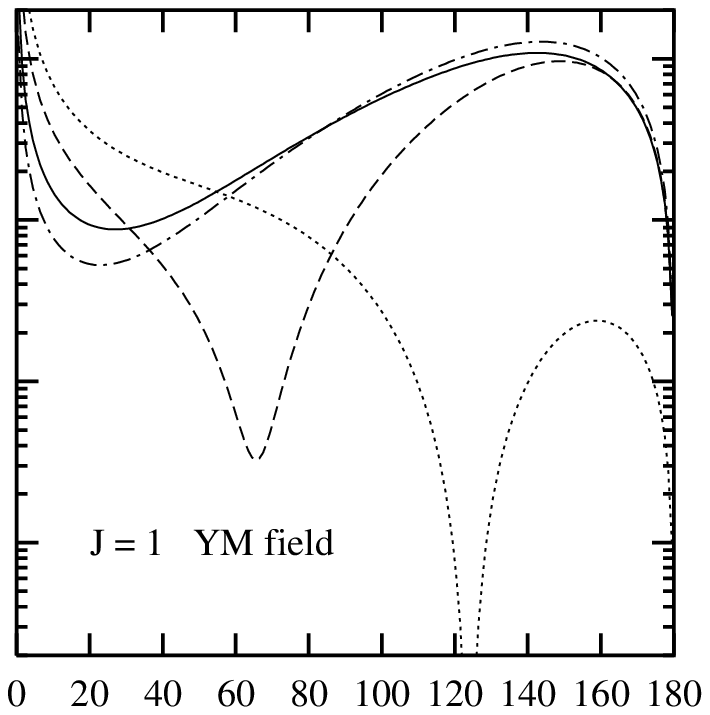,bbllx=4cm,bblly=4.2cm}
\epsfig{file=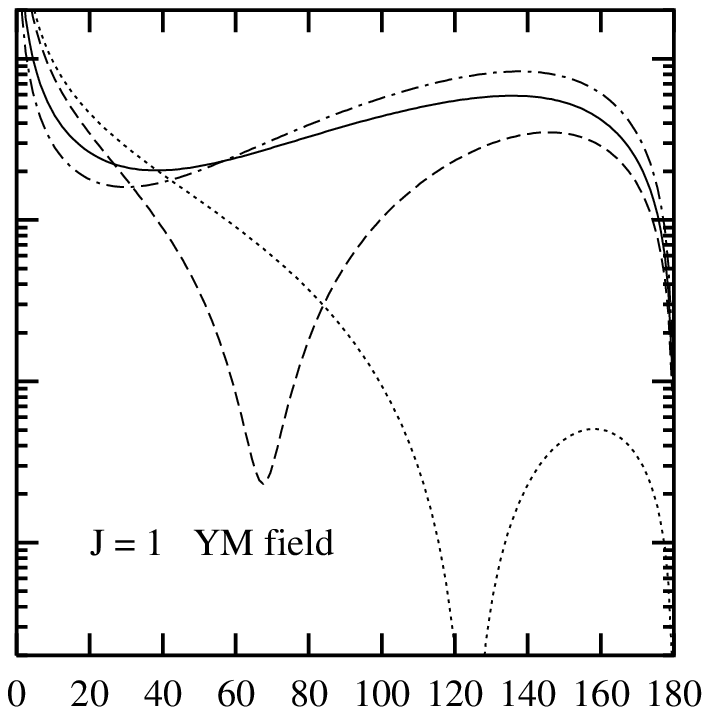,bbllx=-7cm,bblly=4.2cm}
\vskip15mm
\hskip43mm{\large$\theta$}\hskip70mm{\large$\theta$}
\vskip0mm
\caption{
  Angular distribution of the \lk\ production \xs.
  The plotted angle is spanning the directions
  of the incoming electron and the outgoing \lk.
  The collider energy is 1 TeV
  and the \lk\ mass is 400 GeV.
  The coding of the curves is\quad
  solid: $Q=-5/3$;
  dot-dashed: $Q=-4/3$;
  dashed: $Q=-2/3$;
  dotted: $Q=-1/3$.
  Same polarizations as in Fig.~\protect\ref{eny}.
}
\label{ang}
\end{figure}

These very salient features
could well be observed
away from threshold,
where the $u$-channel pole is no longer so dominant.
Of course,
the convolution with the photon energy and polarization spectra
washes out to some extent these prominent features.
But,
as can be gathered from Fig.~\ref{ang},
even so
the angular distributions of the different \lk s
retain their distinctive characteristics.
In this figure,
we have considered 400 GeV \lk s produced at a 1 TeV collider.

To roughly estimate the \ep\ potential
for discriminating the different types of \lk s,
we compare these differential distributions
with a Kolmogorov-Smirnov test.
Say $a$ and $b$ are two different kinds of \lk s
and $a$ is the one which is observed.
The probability that $b$ could be mistaken for $a$
is related to the statistic
\beq
D~=~\sqrt{N_a} ~\sup
\left|
{1\over\sigma_a}\int^\theta\d\theta{\d\sigma_a\over\d\theta}
-
{1\over\sigma_b}\int^\theta\d\theta{\d\sigma_b\over\d\theta}
\right|
\label{eks}\ ,
\eeq
where $\theta$ is the polar angle of the \lk\
and $N_a={\cal L}\sigma_a$ is the number of observed events.

Focusing on the case of Fig.~\ref{ang},
where a 400 GeV \lk\ is studied at a 1 TeV collider,
Tables~\ref{ts} and \ref{tv}
summarize the minimal values of
$$
{\lambda \over e} \quad \sqrt{{\cal L}/{\rm fb}^{-1}}
$$
needed to tell apart
two different types of \lk s
with 99.9\%\ confidence,
{\em i.e.},
setting $D=1.95$ in Eq.~(\ref{eks}).
Assuming each combination of electron and photon polarizations
has accumulated 50 fb$^{-1}$ of data,
some \lk\ types have so different angular distributions
that a coupling as small as $\lambda=.0085e$
is quite sufficient to tell them apart.
Others need as much as $\lambda=.31e$.
Obviously,
these numbers are only valid for this particular choice
of \lk\ mass and collider energy,
and could be improved with a more sophisticated analysis.
Nevertheless,
they should be indicative of what resolving power
an \ep\ scattering experiment can achieve.

\section{Conclusions}

We have studied \lk\ production
in the \ep\ mode of a \lc\ of the next generation.
To perform this analysis,
we have considered all types of scalar and vector \lk s,
whose interactions with leptons and quarks
conserve lepton and baryon number
and are invariant under the \sm\
$SU(3)_c \otimes SU(2)_L \otimes U(1)_Y$
gauge group.
For the vectors we distinguish between \lk s
which couple minimally to photons
and Yang-Mills fields.

The advantage of an \ep\ experiment over,
{\em e.g.},
\pe\ annihilation
is that the \lk s need not be produced in pairs.
Hence higher masses can be explored.
The disadvantage
is that the \ep\ reactions proceed only via the
{\em a priori}
unknown \lk-lepton-quark couplings.

The \sm\ backgrounds can be reduced to almost zero
by simple kinematical cuts,
and the discovery potential is conveniently summarized
by the scaling relation Eq.~(\ref{osc}).
The spin of the \lk s can easily be determined
at threshold
by inverting the polarization of the photon beam.
Moreover,
each \lk\ type
displays very characteristic angular distributions,
some of which having even radiation zeros.
Therefore,
the prospects for discriminating
different \lk s of the same spin
can be greatly enhanced
by studying angular correlations.

All our results have been obtained
with a realistic electron beam polarization
and Compton backscattered photon spectra.

\begin{ack}
It is a pleasure to thank Stan Brodsky and Clem Heusch
for their hospitality at SLAC and UCSC,
where this project was initiated.
Many fruitful discussions with
Sacha Davidson,
Paul Frampton,
Slava Ilyin,
David London and
H\'el\`ene Nadeau
are gratefully acknowledged.
I am also particularly indebted to
David London
for pointing out Refs~\cite{hp,ce,nl,bln,eboli,dg} to me and to
H\'el\`ene Nadeau
for helping me eliminate an error from my calculations.
\end{ack}

\vfill

\newcommand{\rb}[1]{\raisebox{2ex}[-2ex]{$#1$}}

\begin{table}[htb]
$$
\begin{array}{|c|c|c||c|c|c|c|}
\hline
\multicolumn{3}{|c||}{} & \multicolumn{4}{|c|}{J=0}
\\\cline{4-7}
\multicolumn{3}{|c||}
{\begin{picture}(60,0)(0,0)
        \mbox{\boldmath$\displaystyle\Large {\lambda \over e}\sqrt{{\cal
L}/{\rm fb}^{-1}} $}
\end{picture}}
& \multicolumn{2}{|c}{F=0} & \multicolumn{2}{|c|}{F=2}
\\\cline{4-7}
\multicolumn{3}{|c||}{} & Q=-{5\over3} & Q=-{2\over3} & Q=-{4\over3} &
Q=-{1\over3}
\\\hline\hline
& & Q=-{5\over3} & \times & .2 & .7 & .2
\\\cline{3-7}
& \rb{F=0} & Q=-{2\over3} & .8 & \times & .6 & 2.2
\\\cline{2-7}
\rb{J=0} & & Q=-{4\over3} & 1.1 & .3 & \times & .3
\\\cline{3-7}
& \rb{F=2} & Q=-{1\over3} & .4 & 1.0 & .3 & \times
\\\hline
\end{array}
$$
\bigskip
\caption{
Smallest values of
$\lambda/e\protect\sqrt{{\cal L}/{\rm fb}^{-1}}$
which allow discriminating
with 99.9\%\ confidence
two different 400 GeV scalar \lk s
at a 1 TeV collider.
The \lk s listed in the rows
are assumed to produce the data,
whereas those listed in the columns
are tested against this data.
}
\label{ts}
\end{table}

\bigskip

\begin{table}[htb]
\hspace{-1cm}$$
\begin{array}{|c|c|c||c|c|c|c|c|c|c|c|}
\hline
\multicolumn{3}{|c||}{} & \multicolumn{4}{|c|}{J=1} &
\multicolumn{4}{|c|}{J=1\quad\mbox{YM field}}
\\\cline{4-11}
\multicolumn{3}{|c||}
{\begin{picture}(60,0)(0,0)
        \mbox{\boldmath$\displaystyle\Large {\lambda \over e}\sqrt{{\cal
L}/{\rm fb}^{-1}} $}
\end{picture}}
 & \multicolumn{2}{|c}{F=0} & \multicolumn{2}{|c|}{F=2} &
\multicolumn{2}{|c}{F=0} & \multicolumn{2}{|c|}{F=2}
\\\cline{4-11}
\multicolumn{3}{|c||}{} & Q=-{5\over3} & Q=-{2\over3} & Q=-{4\over3} &
Q=-{1\over3} & Q=-{5\over3} & Q=-{2\over3} & Q=-{4\over3} & Q=-{1\over3}
\\\hline\hline
& & Q=-{5\over3} & \times & .3 & .8 & .1 & .3 & .2 & .3 & .1
\\\cline{3-11}
& \rb{F=0} & Q=-{2\over3} & .9 & \times & .7 & .8 & .5 & .5 & .5 & .5
\\\cline{2-11}
\rb{J=1} & & Q=-{4\over3} & 1.2 & .3 & \times & .2 & .6 & .3 & .5 & .1
\\\cline{3-11}
& \rb{F=2} & Q=-{1\over3} & .3 & .5 & .2 & \times & .2 & .2 & .2 & .9
\\\hline
& & Q=-{5\over3} & .2 & .1 & .3 & .07 & \times & .2 & .8 & .06
\\\cline{3-11}
\rb{J=1} & \rb{F=0} & Q=-{2\over3} & .6 & .4 & .6 & .3 & 1.0 & \times & .9 & .3
\\\cline{2-11}
\rb{\mbox{YM}} & & Q=-{4\over3} & .3 & .1 & .3 & .09 & 1.1 & .3 & \times & .08
\\\cline{3-11}
\rb{\mbox{field}} & \rb{F=2} & Q=-{1\over3} & .2 & .4 & .2 & 1.0 & .2 & .2 & .2
& \times
\\\hline
\end{array}
$$
\bigskip
\caption{
Same as Table~\protect\ref{ts}
for the vector \lk s.
Scalar and vector \lk s
can easily be told apart
by flipping the laser polarization at threshold.
}
\label{tv}
\end{table}

\vfill

\clearpage

\begin{figure}[htb]
\centerline{\input eny.tex}
\vspace{-1.5cm}
\centerline{\input pol.tex}
\vspace{-1.5cm}
\centerline{\input ang.tex}
\caption{
  Back-scattered photon energy spectrum,
  polarization and polar angle
  as functions of $x=E_\gamma/E_{\rm beam}$
  for three different combinations of beam polarizations.
}
\label{photon}
\end{figure}

\clearpage

\thispagestyle{empty}

\clearpage

\clearpage

\thispagestyle{empty}


\begin{thebibliography}{88}
\bibitem{etc}{
    see, {\em e.g.}, 
    S. Dimopoulos, J. Ellis,
    {\em Nucl.~Phys.}~{\bf B182} (1981) 505; \\
    O. Shenkar,
    {\em Nucl.~Phys.}~{\bf B206} (1982) 253.
}
\bibitem{comp}{
    see, {\em e.g.}, 
    W. Buchm\"uller,
    {\em Phys.~Lett.}~{\bf B145} (1984) 151; \\
    B. Schrempp, F. Schrempp,
    {\em Phys.~Lett.}~{\bf B153} (1985) 101.
}
\bibitem{gut}{
    see, {\em e.g.}, 
    P.H. Frampton,
    {\em Mod.~Phys.~Lett.}~{\bf A7} (1992) 559.
}
\bibitem{ss}{
    see, {\em e.g.}, 
    J. Hewett, T. Rizzo,
    {\em Phys.~Rep.}~{\bf 183} (1989) 193.
}
\bibitem{brw}{
    W. Buchm\"uller, R. R\"uckl, D. Wyler,
    {\em Phys.~Lett.}~{\bf B191} (1987) 442.
}
\bibitem{ed}{
    J. Bl\"umlein, E. Boos, A. Pukhov,
    {\em Mod.~Phys.~Lett.}~{\bf A9} (1994) 3007
    [hep-ph/9404321].
}
\bibitem{slava}{
    V. Ilyin {\em et al.},
    {\em Phys.~Lett.}~{\bf B351} (1995) 504
    ({\em err.}~{\bf B352} (1995) 500),
    {\em ibid.}~{\bf B356} (1995) 531.
}
\bibitem{sacha}{
    S. Davidson, D. Bailey, B. Campbell,
    {\em Z. Phys.} {\bf C61} (1994) 613
    [hep-ph/9309310]; \\
    M. Leurer,
    {\em Phys.~Rev.}~{\bf D49} (1994) 333
    [hep-ph/9309266],
    {\em ibid.}~{\bf D50} (1994) 536
    [hep-ph/9312341].
}
\bibitem{br}{
    J. Bl\"umlein, R. R\"uckl,
    {\em Phys.~Lett.}~{\bf B304} (1993) 337; \\
    D. Choudhury
    {\em Phys.~Lett.}~{\bf B346} (1995) 291
    [hep-ph/9408250].
}
\bibitem{hp}{
  J.L. Hewett, S. Pakvasa,
  {\em Phys.~Lett.}~{\bf B227} (1989) 178.
}
\bibitem{ce}{
    J.E. Cieza Montalvo, O.J.P. \'Eboli,
    {\em Phys.~Rev.}~{\bf D47} (1993) 837
    [hep-ph/9208242].
}
\bibitem{nl}{
    H. Nadeau, D. London,
    {\em Phys.~Rev.}~{\bf D47} (1993) 3742
    [hep-ph/9303238].
}
\bibitem{bln}{
    G. B\'elanger, D. London and H. Nadeau,
    {\em Phys.~Rev.}~{\bf D49} (1994) 3140
    [hep-ph/9307324].
}
\bibitem{eboli}{
    T.M. Aliev, Kh.A. Mustafaev,
    {\em Sov.~J. Nucl.~Phys.}~{\bf 53} (1991) 482; \\
    O.J.P. \'Eboli {\em et al.},
    {\em Phys.~Lett.}~{\bf B311} (1993) 147
    [hep-ph/9306229]; \\
    M.A. Doncheski, S. Godfrey,
    {\em Phys.~Rev.}~{\bf D49} (1994) 6220
    [hep-ph/9311288].
}
\bibitem{dg}{
    M.A. Doncheski, S. Godfrey,
    {\em Phys.~Rev.}~{\bf D51} (1995) 1040
    [hep-ph/9407317].
}
\bibitem{ginzburg}{
    I.F.~Ginzburg, G.L.~Kotkin, V.G.~Serbo, V.I.~Telnov,
    {\em Nucl.~Instr.~Meth.} {\bf 205} (1983) 47.
}
\end{thebibliography}
\end{document}